\documentclass[11pt,a4paper]{article}
\usepackage{mlmodern}
\usepackage[T1]{fontenc}
\usepackage{verbatim}
\usepackage{float}
\usepackage{amsmath}
\usepackage{amssymb}
\usepackage{graphicx}
\usepackage{setspace}
\usepackage{subfig}
\usepackage[colorlinks=true,allcolors=indigo,backref]
{hyperref}

\usepackage{color}
\usepackage{setspace}
\usepackage{caption}
\newcommand{\mubar}{\bar{\mu}}
\usepackage{xcolor}
\usepackage{braket}
\usepackage{mathrsfs}
\usepackage{fullpage}
\usepackage{cite}
\usepackage{slashed}
\usepackage{epstopdf}
\usepackage[ddmmyyyy]{datetime}
\graphicspath{{./diagrams/}}

\allowdisplaybreaks
\definecolor{indigo}{rgb}{0.0, 0.25, 0.42}
\usepackage{cleveref}
\usepackage{array}

\renewcommand*{\backref}[1]{}
\renewcommand*{\backrefalt}[4]{({%
		\ifcase #1 Not cited.%
		\or Cited on page~#2%
		\else Cited on pages #2%
		\fi%
	})}

\newcommand{\E}{\epsilon}

\newcommand{\nn}{\nonumber \\}
\newcommand{\mhsq}{m_H^2}
\newcommand{\tbf}[1]{{\bf T}^{#1}}
\newcommand{\angspinor}[1]{|#1\rangle}
\newcommand{\sqspinor}[1]{|#1]}
\newcommand{\AsqNLP}[2]{	[\mathcal{A}^2]^{#1}_{#2}|_{\text{NLP}}}

\begin{document}
\begin{titlepage} 
	\begin{flushleft}
		Preprint  
	\end{flushleft}
	
	\begin{flushleft}
		\vspace*{5cm}
		{\Large \bfseries On H+jet production at NLP accuracy}
		\bigskip
		\bigskip
		\medskip \\
		\hrule height 0.05cm
	\end{flushleft}
	\begin{flushleft}
		\vspace{1.0cm}
		\textbf{\large{\sffamily Sourav Pal and Satyajit Seth}} \medskip \\
	\end{flushleft}
	
	\begin{flushleft}
		\textit{Theoretical Physics Division, Physical Research Laboratory,
		}\\
		\textit{Navrangpura, Ahmedabad 380009, India }
	\end{flushleft}
	
	\noindent \textit{E-mail: \href{mailto:sourav@prl.res.in}{sourav@prl.res.in},
		\href{mailto:seth@prl.res.in}{seth@prl.res.in}
		\medskip
	}\\
	
	\noindent \textsc{Abstract}: We present computation of the next-to-leading power corrections for Higgs plus one jet production in a hadron collider via gluon fusion channel. Shifting of spinors in the helicity amplitudes without additional radiation captures the leading next-to-soft radiative behaviour and makes the calculation tractable. We establish the connection between the shifted dipole spinors and the colour ordered radiative amplitudes. We find that next-to-maximal helicity violating amplitudes do not play a role in this correction. Compact analytic expressions of next-to-leading power leading logarithms coming from different helicity configurations are shown.
	
\end{titlepage}

\hrule 
\tableofcontents \vspace{0.5cm}\hrule \vspace{0.4cm}

	
%
\section{Introduction}
Precise experimental data from the Large Hadron Collider and the lack of any persuasive new physics signature demand improvement in the understanding of the Standard Model. Typically in collider environments the strong force dominates over other interactions and that makes the study of theory of Quantum Chromodynamics (QCD) most important. Fixed order corrections by taking into account higher order perturbative terms in the strong coupling constant, and resummation including certain enhanced logarithms to all orders in the perturbation series are the two ways to ameliorate the theoretical accuracy. For all collider processes, one can define a threshold variable that vanishes in the threshold limit. In terms of a generic threshold variable ($\xi$) the differential cross-section takes the following form:  
\begin{align}
	\frac{d\sigma}{d\xi}\,\approx \,\sum_{n=0}^{\infty}\alpha_s^n\left \{\sum_{m=0}^{2n-1}C_{nm}\left (\frac{\log^m\xi}{\xi}\right )_{+} + d_n\delta(\xi)+\sum_{m=0}^{2n-1}D_{nm}\,\log^m\xi\right \}\,.
	\label{eq:gen-threshold}
\end{align}
The first set of logarithms and the delta function are associated with the leading power (LP) approximations, whereas the second set of logarithms appear due to the next-to-leading power (NLP) approximation. The LP terms are well known to originate from the emission of soft and collinear radiation. The seminal works of refs.~\cite{Parisi:1980xd,Curci:1979am,Sterman:1987aj,Catani:1989ne,Catani:1990rp,Gatheral:1983cz,Frenkel:1984pz, Sterman:1981jc} based on diagrammatics helped in devising methods of LP resummation. Later, several alternative methods of LP resummation based on Wilson lines~\cite{Korchemsky:1993xv,Korchemsky:1993uz}, renormalisation group (RG)~\cite{Forte:2002ni} and Soft Collinear Effective Theory (SCET)~\cite{Becher:2006nr,Schwartz:2007ib,Bauer:2008dt, Chiu:2009mg} were developed. A comparative study of different approaches can be found   in~refs.~\cite{Luisoni:2015xha,Becher:2014oda,Campbell:2017hsr}. 

Despite substantial progress made towards understanding the infrared behaviour of the NLP logarithms during the past decade, the universality of such terms is yet to be established. The numerical impacts of NLP logarithms are shown in refs.~\cite{Kramer:1996iq, Ball:2013bra, Bonvini:2014qga, Anastasiou:2015ema, Anastasiou:2016cez, vanBeekveld:2019cks,  vanBeekveld:2021hhv, Ajjath:2021lvg}. Realising the importance of these numerical impacts, several methods to resum NLP logarithms have been formulated over the years~\cite{Grunberg:2009yi,Soar:2009yh, Moch:2009hr,Moch:2009mu,Laenen:2010uz,Laenen:2008gt,deFlorian:2014vta,Presti:2014lqa,Bonocore:2015esa,Bonocore:2016awd,Bonocore:2020xuj,Gervais:2017yxv,Gervais:2017zky,Gervais:2017zdb,Laenen:2020nrt,DelDuca:2017twk,vanBeekveld:2019prq,Bonocore:2014wua,Bahjat-Abbas:2018hpv,Ebert:2018lzn,Boughezal:2018mvf,Boughezal:2019ggi,Bahjat-Abbas:2019fqa,Ajjath:2020ulr,Ajjath:2020sjk,Ajjath:2020lwb,Ahmed:2020caw,Ahmed:2020nci,Ajjath:2021lvg,Kolodrubetz:2016uim,Moult:2016fqy,Feige:2017zci,Beneke:2017ztn,Beneke:2018rbh,Bhattacharya:2018vph,Beneke:2019kgv,Bodwin:2021epw,Moult:2019mog,Beneke:2019oqx,Liu:2019oav,Liu:2020tzd,Boughezal:2016zws,Moult:2017rpl,Chang:2017atu,Moult:2018jjd,Beneke:2018gvs,Ebert:2018gsn,Beneke:2019mua,Moult:2019uhz,Liu:2020ydl,Liu:2020eqe,Wang:2019mym,Beneke:2020ibj,vanBeekveld:2021mxn}. It is essential to investigate NLP logarithms of several processes to better understand the universal nature of the next-to-soft radiation and to come up with a global resummation formulae. The universality of NLP logarithms is already established in case of colour singlet production~\cite{DelDuca:2017twk}, however for coloured particles in the final state there exists no unique resummation formula. 

A prescription has been developed in~\cite{DelDuca:2017twk} for colourless particles and then further extended to final state coloured particles in~\cite{vanBeekveld:2019prq}, in which appropriate shifting of pairs of momenta in the squared non-radiative ({\em i.e.,} without additional radiation) amplitude captures the next-to-soft radiation effects. The expression of the squared non-radiative amplitude does not always have a compact analytical form and therefore shifting the momenta may not always give a simple result. For example, the calculation of NLP terms using the squared amplitudes appears to be very intricate for coloured particles in the final state and due to this reason, only two processes with a single coloured particle in the final state are studied so far at NLP accuracy -- ({\em i}) prompt photon plus jet production~\cite{vanBeekveld:2019prq}, ({\em ii}) W plus jet production~\cite{Sterman:2022lki,Boughezal:2019ggi,vanBeekveld:2023gio}. The scarcity of results for coloured final state particles due to the complexity in such calculations clearly demands an improvement on the existing technique.  

In this endeavour, we study the effect of next-to-soft gluon radiation on Higgs production via gluon fusion in association with a final state hard jet by crafting spinor helicity amplitudes. We consider the heavy top mass limit throughout. Instead of shifting momenta in the squared amplitude, we shift the spinors of the non-radiative helicity amplitudes to capture next-to-soft radiation effects and that essentially makes the calculation lucid and tractable. We start with the soft and next-to-soft theorems developed in~\cite{Casali:2014xpa, Luo:2014wea} and show that pairwise shifting of spinors at the non-radiative amplitudes can be realised as next-to-soft emissions from those amplitudes. We find that colour dipoles with shifted spinors directly correspond to the colour ordered radiative amplitudes in the next-to-soft limit. The next-to-soft amplitudes thus obtained are compact in nature. In addition, it reveals that the next-to-maximal helicity violating (NMHV) amplitudes never contribute at NLP accuracy for the case at hand. In order to obtain NLP logarithms, we integrate squared helicity amplitudes over the unresolved parton phase space and present the analytic results for different helicity configurations. Singularities that arise at the LP and NLP stages get exactly cancelled while contributions from the virtual emission and mass factorisation are included.   

The structure of our paper is as follows. In section~\ref{sec:shifts}, we review the soft and next-to-soft theorems in terms of spinor shifts. After detailing the shifts, in section~\ref{sec:LPplusNLP} we apply them to calculate different colour ordered helicity amplitudes. Squaring these amplitudes, we perform the phase space integration over the unresolved phase space in section~\ref{sec:phasespace} to obtain the NLP logarithms. Finally, we summarise our findings with an outlook in section~\ref{sec:conclu}. Throughout this study, we have used a combination of in-house routines based on \texttt{QGRAF}~\cite{Nogueira:1991ex}, \texttt{FORM}~\cite{Vermaseren:2000nd} and \texttt{Mathematica}~\cite{Mathematica} to calculate all helicity amplitudes and to perform the phase space integration.   

\section{Soft and next-to-soft corrections}
\label{sec:shifts}
In this section, we briefly review the soft and next-to-soft theorems in terms of colour ordered scattering amplitudes. Any colour ordered scattering amplitude involving $n$ particles (quarks and gluons) with specific helicities can be represented as, 
\begin{align}
\mathcal{A}\,=\, \mathcal{A}_{n}\left(\left\{ |1\rangle,|1]\right\} ,\ldots,\left\{ |n\rangle,|n]\right\} \right)\,,
\end{align}
where $ \angspinor{i} $ and $ \sqspinor{i} $ denote the holomorphic and anti-holomorphic spinors associated with the particle $i$ carrying momentum $p_i$. Let us now consider that a gluon $s$ with momenta $p_s$ and helicity \lq $+$\rq\,  is being emitted from this scattering process. Scaling the momentum of the radiated gluon $p_s\to\lambda\,p_s$, the scattering amplitude for $ n+1 $ particle can be expressed in powers of $ \lambda$ as written here under~\cite{Luo:2014wea,Casali:2014xpa}, 
\begin{equation}
	\mathcal{A}_{n+1}\bigg(\left\{ \lambda\angspinor{s},\sqspinor{s}\right\} ,\left\{ \angspinor{1},\sqspinor{1}\right\} ,\ldots,\left\{ \angspinor{n},\sqspinor{n}\right\} \bigg)=\left(S^{(0)}+S^{(1)}\right)\mathcal{A}_{n}\bigg(\left\{ \angspinor{1},\sqspinor{1}\right\} ,\ldots,\left\{ \angspinor{n},\sqspinor{n}\right\} \bigg)\,.
	\label{eq:factorization}
\end{equation}
Here $ S^{(0)} $ and $ S^{(1)} $ denote LP and NLP terms that are of $\mathcal{O}(1/\lambda^2)$ and $\mathcal{O}(1/\lambda)$ respectively, and are given by, 
\begin{align}
	S^{(0)} &\, =\,\frac{\braket{n1}}{\braket{s1}\braket{ns}} \,, \nn 
	S^{(1)}&\, =\,\frac{1}{\braket{s1}}\sqspinor{s}\frac{\partial}{\partial\sqspinor{1}}-\frac{1}{\braket{sn}}\sqspinor{s}\frac{\partial}{\partial\sqspinor{n}}\,.
	\label{eq:leading}
\end{align}
In order to obtain the above formulae, holomorphic soft limit~\cite{Casali:2014xpa,Luo:2014wea} is being used {\em i.e.,} 
\begin{align}
\angspinor{s}\rightarrow \lambda\, \angspinor{s}\,,\quad \sqspinor{s}\rightarrow \sqspinor{s} \,,
 \end{align} 
under the BCFW~\cite{Britto:2004ap,Britto:2005fq} deformation of $s$ and $n$ pair, while particles $1$ and $s$ always form a three particle amplitude involving the on-shell cut propagator that carries complex momentum. With the help of eq.~\eqref{eq:leading}, the colour ordered amplitude of eq.~\eqref{eq:factorization} can be rewritten as,  
\begin{align}
\mathcal{A}_{n+1}^{\text{LP+NLP}}\bigg(\left\{ \lambda\angspinor{s},\sqspinor{s}\right\} ,\left\{ \angspinor{1},\sqspinor{1}\right\} ,\ldots,\left\{ \angspinor{n},\sqspinor{n}\right\}  \bigg)& \,  = \, \frac{1}{\lambda^2} \frac{\braket{1n}}{\braket{1s}\braket{ns}}  \mathcal{A}_n\bigg(\left \{\angspinor{1}, \sqspinor{1^\prime} \right \},  \ldots,\left \{ \angspinor{n}, \sqspinor{n^\prime} \right \}\bigg) \,,
\label{eq:LPplusNLP}
\end{align}
where 
	\begin{align}
	\sqspinor{1^\prime} & \,=\, \sqspinor{1}+\Delta_s^{(1,n)}\sqspinor{s}\nonumber \,,\\
	\sqspinor{n^\prime} & \,=\,\sqspinor{n}+\Delta_s^{(n,1)}\sqspinor{s} \,,
	\label{eq:gen-shifts}
\end{align}
and, 
\begin{equation}
	\Delta_s^{(i,j)}=\lambda \frac{\braket{js}}{\braket{ji}}\,.
	\label{eq:mom-ratio}
\end{equation}
This form of eq.~\eqref{eq:LPplusNLP} signifies that the leading and subleading behaviour of the amplitude can be obtained in terms of simple shifts in the spinors of tree amplitudes. Note that the emitted soft gluon is placed in between the $1$ and $n$ particles in the colour ordered amplitudes and forms a colour dipole $\mathcal{D}_{1n}$. Such colour dipole structures play an important role in understanding the IR singularities of  scattering amplitudes~\cite{Catani:1996vz,Gardi:2009qi,Becher:2009cu}. 

Emission of soft gluon with \lq $-$\rq \, helicity can be treated analogously by taking anti-holomorphic soft gluon limit and interchanging angle and square spinors. Equipped with these formulae, we now move on to calculate the LP and NLP amplitudes for Higgs plus one jet production in the gluon fusion channel. 

\section{LP and NLP amplitudes for $ gg \rightarrow Hg$}
\label{sec:LPplusNLP}
The most dominant mechanism for Higgs boson production at the LHC is via the gluon fusion channel. In this section we first reproduce all independent helicity amplitudes for Higgs plus one jet production via gluon fusion with(out) one extra gluon emission. Then we obtain NLP amplitudes by -- ({\em i}) taking soft gluon limit on $gg\to Hgg$ amplitudes, ({\em ii}) shifting spinors in $gg\to Hg$ amplitudes. Both ways lead to the exactly same results.  Finally we discuss that for Higgs plus one jet production NMHV amplitudes do not contribute to the NLP threshold corrections.    

\subsection{Higgs-gluon amplitudes}
The Standard Model of particle physics forbids gluons to interact with Higgs at the tree level, however they can interact via a massive quark loop. As the top quark is the heaviest among massive quarks, the coupling of Higgs with gluons is dominated via a top quark loop. In the large top mass limit $ m_t\rightarrow\infty $, we can integrate out the heavy top quark effect to obtain an effective Lagrangian as follows~\cite{Wilczek:1977zn, Shifman:1978zn}, 
\begin{align}
	\mathcal{L}_{\,\text{eff}}\,=\, -\frac{1}{4}\, G\, \,H \,\text{Tr} (F_{\mu \nu }^a F^{\mu \nu,a})\,,
\end{align}
where $ F_{\mu \nu}^a $ is the QCD field strength tensor. 
The effective coupling is given at lowest order by $ G=\alpha_{s}/3\pi v $, where $v$ is the vacuum expectation value of the Higgs field and $ \alpha_s $ is the strong coupling constant.
The general form of an amplitude consisting of one Higgs boson and $ n $-gluons can be represented as, 
\begin{equation}
	\mathcal{A}_n (p_i,h_i,c_i)\,=\, i\,\left (\frac{\alpha_{s}}{6\pi v}\right )g_s^{n-2} \sum_{\sigma \in \mathcal{S}_{n'}} \text{Tr} \left( \tbf{c_1}\tbf{c_2} \ldots  \tbf{c_n} \right )\mathcal{A}_n^{\{c_i\}} \left (h_1\,h_2\,h_3 \ldots \,{h_n };H \right)\,. 
	\label{eq:gen-ng}
\end{equation}
Here $ \mathcal{S}_{n'} $ represents the set of all $ (n-1)! $ non-cycling permutations of $ 1,2,\ldots,n $. $ \tbf{c_i} $ denote the SU(3) colour matrix in the fundamental representation and the are normalized as, $ \text{Tr} (\tbf{c_1},\tbf{c_2})= \delta^{c_1 c_2} $. For brevity, we avoid writing $H$ explicitly in $\mathcal{A}_n^{\{c_i\}}$ in the rest of this paper.

The leading order process for Higgs plus one gluon production can be written as,
\begin{align}
	g (p_1)+ g(p_2) \rightarrow H(-p_3) +g (-p_4)\,.
	\label{eq:momassign}
\end{align}
There are two independent colour ordered helicity amplitudes for this process as given below, 
\begin{align}
	\mathcal{A}^{124}_{+++} & =\frac{m_{H}^{4}}{\braket{12}\braket{24}\braket{41}}\,, \qquad 
	\mathcal{A}^{124}_{-++}  =\frac{[24]^{3}}{[12][14]}\,,
	\label{eq:LO}
\end{align}
and amplitudes for all other helicity configurations can be constructed using these two. 

Now, we consider that a gluon with momenta $ p_5 $ is being emitted from the leading order process, {\em i.e.,}  
\begin{align}
	g (p_1)+g(p_2) \rightarrow H(-p_3)+g(-p_4)+g(-p_5) \, .
\end{align}
For this process, there are only three independent helicity amplitudes and remaining helicity configurations can be obtained by switching external momenta and spinors. These three independent helicity amplitudes containing Higgs plus four gluons are given by, 
\begin{align}
	\mathcal{A}^{1245}_{++++}\,& =\frac{m_{H}^4}  {\braket{1 2} \braket{2 4} 
		\braket{ 45} \braket{5 1}} \,,\nn
	\mathcal{A}^{1245}_{-+++}\,& =\frac{\langle 1|4+5|2 ]^3}{\langle4|1|2] \braket{15} \braket{45} s_{145}}+\frac{[25] [45] \langle 1|4+5|2 ]^2}{\langle4|1|2] s_{15} s_{145}} +\frac{[24] \langle 1|2+4|5 ]^2}{\braket{24} s_{12}  s_{124}}
\nn& \quad +\frac{[25] \langle 1|2+4|5 ]^2}{ \braket{14} \braket{24} [15] s_{12}} -\frac{[25]^2 \langle 1|2+5|4 ]^2}{s_{12} s_{15} s_{125}} \,,\nn 
	\mathcal{A}^{1245}_{--++}\,&=-{\langle 1 2\rangle^4\over \langle 12 \rangle 
		\langle 24 \rangle \langle 45\rangle \langle 51\rangle}
	- {[45]^4 \over [1 2] [24] [45] [51]}\,.
\end{align}	
Here $s_{ij}=(p_i+p_j)^2$ and $s_{ijk}=(p_i+p_j+p_k)^2$. These amplitudes were calculated for the first time in~\cite{Kauffman:1996ix}. 
Following eq.~\eqref{eq:gen-ng}, we can write the full amplitude for a given helicity configuration as, 
\begin{align}
	& \mathcal{A}(\{p_i,h_i,c_i\})\,\nn& =\, i \, \left (\frac{\alpha_{s}}{6\pi v}\right )\, g_s^2 \Bigg[ \left \{\text{Tr} \left (\tbf{c_1}\tbf{c_2}\tbf{c_4}\tbf{c_5}\right )+ \text{Tr}\left (\tbf{c_1}\tbf{c_5}\tbf{c_4}\tbf{c_2}\right )\right \} \mathcal{A}^{1245}_{h_1h_2h_4h_5} \nn & 
	+\left \{\text{Tr} \left (\tbf{c_1}\tbf{c_4}\tbf{c_5}\tbf{c_2}\right )+ \text{Tr}\left (\tbf{c_1}\tbf{c_2}\tbf{c_5}\tbf{c_4}\right )\right \} \mathcal{A}^{1452}_{h_1h_2h_4h_5} \nn &
	+\left \{\text{Tr} \left (\tbf{c_1}\tbf{c_5}\tbf{c_2}\tbf{c_4}\right )+ \text{Tr}\left (\tbf{c_1}\tbf{c_4}\tbf{c_2}\tbf{c_5}\right )\right \} \mathcal{A}^{1524}_{h_1h_2h_4h_5} \Bigg] \,.
	\label{eq:gen4gamp}
\end{align}
Squaring the above equation and summing over colours, we obtain the expression of squared amplitude as, 
\begin{align}
	\sum_{\text{colours}}|\mathcal{A}(\{p_i,h_i,c_i\})|^2\,& =\,	\bigg[\left (\frac{\alpha_{s}}{6\pi v}\right )\, g_s^2\bigg]^2 (N^2-1) \bigg \{ 2\, N^2 \left ( |\mathcal{A}^{1245}|^2+|\mathcal{A}^{1452}|^2+|\mathcal{A}^{1524}|^2\right) \nn
	&  -4 \frac{(N^2-3)}{N^2} \big|\mathcal{A}^{1245}+\mathcal{A}^{1452}+\mathcal{A}^{1524}\big|^2 \bigg\}\,.
	\label{eq:genAsq}
\end{align}
Here, for simplicity, we have suppressed the labels that represent helicity configurations. Due to the dual Ward identity~\cite{Dixon:2004za,Kauffman:1996ix}, the term in the second line of the above equation vanishes and we are left with only the first term. 

\subsection{Spinor shifts and colour dipoles}
In order to obtain NLP amplitudes for the Higgs plus two gluon production process, one needs to expand the $gg\to Hgg$ helicity amplitudes in the powers of the soft momentum keeping the sub-leading contributions. In parallel, following the arguments presented in section~\ref{sec:shifts}, we can get NLP amplitudes using the shifts in the spinors of $gg\to Hg$ amplitudes. We start our calculation by noting the fact that the gluon with momentum $ p_5 $ be emitted from any of the three gluons present at the leading order and as discussed in the previous section, the emission of a soft gluon always engenders shifts in two adjacent spinors present in the colour ordered non-radiative Born amplitudes. 

In case of emission of a next-to-soft gluon from Higgs plus $ n $ gluon amplitudes, a total $ ^nC_2= n(n-1)/2 $ number of colour dipoles can be formed. Therefore, for amplitudes consisting of Higgs plus three gluons, three dipoles are generated due to the emission of a next-to-soft gluon and NLP amplitudes can be realised by shifting appropriate spinors depending on the helicity of the emitted gluon. For a \lq $+$\rq \, gluon emisison from the dipole $\mathcal{D}_{14}$ made up of momenta $p_1$ and $p_4$, the LP+NLP amplitude  can be expressed as,
\begin{align}
	\mathcal{A}^{1245}_{h_1 h_2 h_4 +} \, =\, \frac{\braket{14}}{\braket{15}\braket{45}}\mathcal{A}^{1^\prime2\,4^\prime}_{h_1 h_2 h_4} \,,
	\label{eq:gen-shift14}
\end{align}
where $\mathcal{A}^{1^\prime2\,4^\prime}_{h_1 h_2 h_4}$ denotes that the \sqspinor{1} and \sqspinor{4} spinors are shifted in the colour ordered leading amplitude obeying eq.~(\ref{eq:gen-shifts}). Similar contributions coming from the dipoles $\mathcal{D}_{24}$ and $\mathcal{D}_{12}$ can be written as, 
\begin{align}
	\mathcal{A}^{1452} _{h_1h_2h_4+} \, =\, \frac{\braket{24}}{\braket{25}\braket{54}}
	\mathcal{A}^{1\,2'4'}_ {h_1h_2h_4}\,,
	\label{eq:gen-shift24}
\end{align}
and
\begin{align}
	\mathcal{A}^{1524} _{h_1h_2h_4+} \, =\, \frac{\braket{12}}{\braket{15}\braket{52}}\mathcal{A}^{1'2'\,4}_{h_1h_2h_4} \,.
	\label{eq:gen-shift12}
\end{align}
So the full amplitude of eq.~\eqref{eq:gen4gamp} can now be rewritten using eqs.~\eqref{eq:gen-shift14}~--~\eqref{eq:gen-shift12} as, 
\begin{align}
		& \mathcal{A}_{h_1h_2h_4+}|_{\,\text{LP+NLP}}\,=\, i \, \left (\frac{\alpha_{s}}{6\pi v}\right )\, g^2 \nn \times &\Bigg[ \left \{\text{Tr} \left (\tbf{C_1}\tbf{C_2}\tbf{C_4}\tbf{C_5}\right )+ \text{Tr}\left (\tbf{C_1}\tbf{C_5}\tbf{C_4}\tbf{C_2}\right )\right \} \frac{\braket{14}}{\braket{15}\braket{45}}\mathcal{A}^{1^\prime2\,4^\prime}_{h_1 h_2 h_4} \nn & 
		-\left \{\text{Tr} \left (\tbf{C_1}\tbf{C_4}\tbf{C_5}\tbf{C_2}\right )+ \text{Tr}\left (\tbf{C_1}\tbf{C_2}\tbf{C_5}\tbf{C_4}\right )\right \} \frac{\braket{24}}{\braket{25}\braket{45}} \mathcal{A}^{1\,2'4'}_ {h_1h_2h_4} \nn &
		-\left \{\text{Tr} \left (\tbf{C_1}\tbf{C_5}\tbf{C_2}\tbf{C_4}\right )+ \text{Tr}\left (\tbf{C_1}\tbf{C_4}\tbf{C_2}\tbf{C_5}\right )\right \} \frac{\braket{12}}{\braket{15}\braket{25}}\mathcal{A}^{1'2'\,4}_{h_1h_2h_4} \Bigg]\,. 
		\label{eq:gen4gLPNLP}
\end{align}
To derive the above equation, we have used the reflection identity~\cite{Dixon:2004za} that applies for Higgs plus $n$-gluon amplitudes. 
This equation is one of the central results of this paper which identifies the direct correspondence of colour ordered amplitudes in the next-to-soft limit to the non-radiative colour ordered Born amplitudes with shifted spinors. Shift in each non-radiative spinor pair represents one colour ordered radiative amplitude. The validity of this formula relies only on the cyclic and antisymmetric properties of Higgs plus gluon amplitudes. Thus, this formula is applicable to any process that satisfy such properties, namely pure gluon amplitudes in Yang-Mills theories or gluons with a quark-antiquark pair in QCD.     

\subsection{NLP Amplitudes: Absence of NMHV contribution}
As evident from the discussion in the previous section, colour ordered LP amplitudes always appear as a product of Born amplitudes and the corresponding Eikonal factors such as,
\begin{align}
	\left.\mathcal{A}^{1245} _{h_1h_2h_4+}\right|_{\,\rm LP}\,& =\, \frac{\braket{14}}{\braket{15}\braket{45}} \mathcal{A}^{124} _{h_1h_2h_4} \,,\nn
\left.\mathcal{A}^{1245} _{h_1h_2h_4-}\right|_{\,\rm LP}\,& =\, \frac{[14]}{[15][45]} \mathcal{A}^{124} _{h_1h_2h_4} \,.
\end{align} 
In this section we provide the details of NLP amplitudes for different helicity configurations. For Higgs plus four gluon amplitudes, there are altogether $ 2^4=16 $ helicity configurations possible. Out of these sixteen helicity amplitudes, one needs to calculate only eight, as the remaining conjugate configurations can easily be obtained by flipping the helicity of all the external gluons. As discussed earlier, NLP amplitudes can be calculated considering  emission of both \lq $+$\rq \, and \lq $-$\rq \, helicity gluons from all possible Born amplitudes. In doing so, we find that the NMHV amplitudes do not add to the NLP contribution. We illustrate this by a simple example. Let us consider emission of a \lq $+$\rq \, helicity gluon out of the $\mathcal{A}^{124}_{+--}$ amplitude, which following eq.~\eqref{eq:LO} can be presented as, 
\begin{align}
	\mathcal{A}^{124}_{+--}  =-\frac{\braket{24}^{3}}{\braket{12}\braket{14}}\,.
\end{align}
We have already seen in the previous subsection that emission of a \lq $+$\rq \, helicity gluon always demands anti-holomorphic spinors to be shifted. However there are no square spinors present in the above equation and therefore $ \mathcal{A}_{+--+}^{1245}|_{\text{NLP}}$ vanishes. It is also straight forward to check that applying $ S^{(1)} $ of eq. (\ref{eq:leading}) on $ \mathcal{A}^{124}_{+--} $ gives zero. The reason behind this vanishing of NLP amplitude for NHMV amplitudes can furthermore be argued by invoking the soft Higgs limit. Due to the momentum conservation, one can choose not to bring Higgs momentum explicitly in the expressions of NLP amplitudes and in the soft Higgs limit these amplitudes essentially behave as pure gluon NMHV amplitudes which were shown to be non-contributing to NLP in~\cite{Luo:2014wea}.   
\begin{table}[t]
\begin{center}	
	\begin{tabular} {|c|c|c|}
		\hline 
		Born & Helicity of extra emission & NLP \\	
		\hline 
		$\mathcal{A}_{+++}$ & + & $\left .\mathcal{A}_{++++}\right|_{\text{NLP}}$ \\ 
		& - & $\left .\mathcal{A}_{+++-}\right|_{\text{NLP}}$ \\
		\hline 
		$\mathcal{A}_{-++}$	& + & $\left .\mathcal{A}_{-+++}\right|_{\text{NLP}}$ \\
		& - & 0 \\
		\hline 
		$\mathcal{A}_{+-+}$	& + & $\left .\mathcal{A}_{+-++}\right|_{\text{NLP}}$ \\
		& - & $0$ \\
		\hline 
		$\mathcal{A}_{++-}$	& + & $\left .\mathcal{A}_{++-+}\right|_{\text{NLP}}$\\
		& - & 0 \\
		\hline 
	\end{tabular}
\caption{A set of eight NLP amplitudes constructed from the Born amplitudes is given. Flipping helicities of all the particles provide the remaining eight amplitudes among which three NMHV configurations become zero again. We do not mention any explicit colour ordering here as this feature stands true irrespective of that choice. \label{tab:All-helicity}}
\end{center}
\end{table}
Among sixteen Higgs plus four gluon helicity amplitudes, six NMHV amplitudes vanish and we are left with ten non-zero helicity amplitudes at NLP. Out of these ten, we need to calculate only five, as the remaining five helicity configurations can readily be obtained by flipping helicities of all external gluons. Table~\ref{tab:All-helicity} shows emissions from the Born amplitudes and lists down those five different non-zero amplitudes. 
Their expressions including three different colour orderings for each of them are given below. 

\subsection*{1. ${++++} $}
\begin{align}
	\left.\mathcal{A}^{1245}_{++++}\right|_{\rm NLP} &\,  =\,\frac{\braket{14}}{\braket{15}\braket{45}}\,\frac{2\,(s_{15}+s_{25}+s_{45})}{(s_{12}+s_{14}+s_{24})}\mathcal{\,A}^{124}_{+++} \,,\nn
\left.\mathcal{A}^{1524}_{++++}\right|_{\rm NLP}\,& =-\frac{\braket{12}}{\braket{15}\braket{25}}	\frac{2\,(s_{15}+s_{25}+s_{45})}{(s_{12}+s_{14}+s_{24})}\mathcal{\,A}^{124}_{+++}\,, \nn 
	\left.\mathcal{A}^{1452}_{++++}\right|_{\rm NLP}\,&=\, -\frac{\braket{24}}{\braket{45}\braket{25}}	\frac{2\,(s_{15}+s_{25}+s_{45})}{(s_{12}+s_{14}+s_{24})}\mathcal{\,A}^{124}_{+++}\,. 
\label{eq:ppppNLPall}
\end{align}
\subsection*{ 2. $ -+++ $}
\begin{align}
	\left.\mathcal{A}^{1245}_{-+++}\right|_{\rm NLP}\,& =\, \frac{\braket{14}}{\braket{15}\braket{45}} \bigg(\frac{3 \braket{15} [25]}{\braket{14} [24]}-\frac{\braket{45} [25]}{\braket{14} [12]}-\frac{s_{15}}{s_{14}}-\frac{s_{45}}{s_{14}}\bigg) \mathcal{\,A}^{124}_{-++}\,, \nn
\left.\mathcal{A}^{1524}_{-+++}\right|_{\rm NLP}\,& =-\frac{\braket{12}}{\braket{15}\braket{25}} \bigg ( -\frac{\langle 2 5 \rangle [45]}{\langle 1 2 \rangle [14]}-\frac{3 \langle 1 5 \rangle [45]}{\langle 1 2 \rangle [24]}-\frac{s_{15}}{s_{12}}-\frac{s_{25}}{s_{12}} \bigg)\,\mathcal{\,A}^{124}_{-++}\,, \nn 	
\left.\mathcal{A}^{1452}_{-+++}\right|_{\rm NLP}\,&=\, -\frac{\braket{24}}{\braket{45}\braket{25}}\bigg(\frac{\braket{45} [15]}{\braket{24} [12]}-\frac{\braket{25} [15]}{\braket{24} [14]}+\frac{3 s_{25}}{s_{24}}+\frac{3 s_{45}}{s_{24}}\bigg) \mathcal{\,A}^{124}_{-++}\,.	
\label{eq:mpppNLPall}
\end{align}
\subsection*{3. $ ++-+ $}
\begin{align}
\left. \mathcal{A}^{1245}_{++-+}\right|_{\text{NLP}}\,&=\,\left.\mathcal{A}^{1245}_{-+++}\right|_{\rm NLP} \,\{1\leftrightarrow4\}\,, \nn
\left.\mathcal{A}^{1524}_{++-+}\right|_{\rm NLP}\,& =\,\left.\mathcal{A}^{1452}_{-+++}\right|_{\rm NLP} \,\{1\leftrightarrow4\} \,,\nn 
\left .\mathcal{A}^{1452}_{++-+}\right|_{\text{NLP}} \,&=\left.\mathcal{A}^{1524}_{-+++}\right|_{\rm NLP}\, \{1\leftrightarrow4\}\, .
\label{eq:ppmpallNLP}
\end{align}
\subsection*{4. $+-++$}
\begin{align}
	\left.\mathcal{A}^{1245}_{+-++}\right|_{\text{NLP}}\,&=\,\left.\mathcal{A}^{1452}_{-+++}\right|_{\rm NLP} \,\{1\leftrightarrow2\} \,, \nn 
\left.\mathcal{A}^{1524}_{+-++}\right|_{\text{NLP}}\,& =\,\left.\mathcal{A}^{1524}_{-+++}\right|_{\rm NLP} \,\{1\leftrightarrow2\} \,,\nn
\left.\mathcal{A}^{1452}_{+-++}\right|_{\text{NLP}}\,&=\,\left.\mathcal{A}^{1245}_{-+++}\right|_{\rm NLP} \,\{1\leftrightarrow2\}\,.
\label{eq:pmmpallNLP}	
\end{align}	
\subsection*{5. $+++-$}
\begin{align}
	\left.\mathcal{A}^{1245}_{+++-}\right|_{\text{NLP}}\,&=\, \frac{[14]}{[15][45]} \bigg ( -\frac{\braket{25} [45]}{\braket{12} [14]}-\frac{\braket{25} [15]}{\braket{24} [14]}-\frac{s_{15}}{s_{14}}-\frac{s_{45}}{s_{14}}\nn & \qquad \qquad +\frac{2\,(s_{15}+s_{25}+s_{45})}{(s_{12}+s_{14}+s_{24})} \bigg) \mathcal{A}^{124} _{+++} \,,\nn 
	\left.\mathcal{A}^{1524}_{+++-}\right|_{\text{NLP}}\,& =-\frac{[12]}{[15][25]}\,	\bigg (\frac{\braket{45} [15]}{\braket{24} [12]}-\frac{\braket{45} [25]}{\braket{14} [12]}-\frac{s_{15}}{s_{12}}-\frac{s_{25}}{s_{12}}
\nn & \qquad \qquad +\frac{2\,(s_{15}+s_{25}+s_{45})}{(s_{12}+s_{14}+s_{24})}\bigg) \mathcal{A}^{124} _{+++}\,, \nn
	\left.\mathcal{A}^{1452}_{+++-}\right|_{\text{NLP}}\,&=\, -\frac{[24]}{[25][45]} \bigg (
\frac{\braket{15} [45]}{\braket{12} [24]}-\frac{\braket{15} [25]}{\braket{14} [24]}-\frac{s_{25}}{s_{24}}-\frac{s_{45}}{s_{24}} 
\nn & \qquad \qquad +\frac{2\,(s_{15}+s_{25}+s_{45})}{(s_{12}+s_{14}+s_{24})}
\bigg )\, \mathcal{A}^{124} _{+++} \,.
\label{eq:pppmNLPall}
\end{align}

\section{NLP logarithms}
\label{sec:phasespace}
The obvious next step to obtain the NLP threshold logarithms is to perform phase-space integrations over the squared amplitudes at NLP and we discuss that in the following two subsections. 

\subsection{Squared amplitudes at NLP}
The amplitude for a process carrying a soft gluon radiation can be written as a sum of LP and NLP amplitudes such as, 
\begin{align}
	\mathcal{A}\,=\, \mathcal{A}_\text{LP}+\mathcal{A}_\text{NLP} \,.
\end{align}
Squaring the amplitude gives,
\begin{align}
	\mathcal{A}^2\,=\,\mathcal{A}_\text{LP}^2+2\text{Re} \left (\mathcal{A}_\text{NLP}\mathcal{A}_{\text{LP}}^\dagger \right ) \,,
\end{align}
where the term $\mathcal{A}_\text{NLP}^2$ is being neglected as it starts contributing at the next-to-next-to leading power. The first and the second terms represent LP and NLP contributions respectively, and we denote the NLP contribution as $ \AsqNLP{\,}{\,} $ hereafter. Using eq.~\eqref{eq:genAsq} 
 we obtain the squared NLP amplitude for a fixed helicity as, 
\begin{align}
\sum_{\text{colours}}\AsqNLP{}\,& =\,	\bigg[\left (\frac{\alpha_{s}}{6\pi v}\right )\, g^2\bigg]^2 2\, N^2 (N^2-1) \bigg \{\AsqNLP{1245}{}+\AsqNLP{1452}{}+\AsqNLP{1524}{}\bigg\} \,,
\label{eq:genNLPsq}
\end{align}
where $ N $ is the dimensionality of the $ \text{SU} (N) $ colour and it takes the value $ N=3 $ for QCD.  

Using eqs.~\eqref{eq:ppppNLPall} --~\eqref{eq:pppmNLPall} we get the squared NLP amplitudes of the following five helicity configurations, 
\begin{enumerate}
\item $ ++++ $	
\begin{align}
	\AsqNLP{1245}{++++}&\, =\,4 \left(\frac{s_{14} s_{25}}{s_{15} s_{45}}+\frac{s_{14}}{s_{15}}+\frac{s_{14}}{s_{45}}\right) \frac{1}{(s_{12}+s_{14}+s_{24})} \mathcal{\,A}^2_{+++} \,, \nn 
\AsqNLP{1524}{++++}&\,=\,\AsqNLP{1245}{++++}\, \{2\leftrightarrow4\}\,, \nn 
\AsqNLP{1452}{++++}&\,=\,\AsqNLP{1245}{++++}\, \{1\leftrightarrow2\} \,.
\label{eq:ppppNLPsq}
\end{align}
\item $ -+++ $
\begin{align}
\AsqNLP{1245}{-+++}&\,=\, \bigg(-\frac{3\, s_{12}}{s_{15} s_{24}}-\frac{3}{ s_{15}}+\frac{1}{s_{45}}+\frac{s_{24}}{s_{12} s_{45}}-\frac{s_{14} s_{25}}{ s_{12} s_{15} s_{45}}+\frac{3\, s_{14} s_{25}}{ s_{15} s_{24} s_{45}} \bigg) \, \mathcal{A}^2 _{-++} \,, \nn
\AsqNLP{1524} {-+++}&\,=\, \AsqNLP{1245}{-+++}\, \{2\leftrightarrow4\} \,, \nn
\AsqNLP{1452} {-+++} & \,=\, \bigg(\frac{s_{12}}{s_{14} s_{25}}+\frac{5}{s_{25}}+\frac{5}{s_{45}}+\frac{s_{14}}{s_{12} s_{45}}-\frac{s_{15} s_{24}}{s_{12} s_{25} s_{45}}-\frac{s_{15} s_{24}}{s_{14} s_{25} s_{45}}\bigg)  
\mathcal{A}^2 _{-++} \,.
\end{align}
\item $ ++-+ $
\begin{align}
\AsqNLP{1245}{++-+} \,=\, \AsqNLP{1245}{-+++}\,\{1\leftrightarrow4\} \,,\nn 
\AsqNLP{1524}{++-+} \,=\, \AsqNLP{1452}{-+++}\,\{1\leftrightarrow4\} \,, \nn 
\AsqNLP{1452}{++-+} \,=\, \AsqNLP{1524}{-+++}\,\{1\leftrightarrow4\}  \,.
\end{align}
\item $ +-++ $
\begin{align}
\AsqNLP{1245}{+-++}\,=\, \AsqNLP{1452}{-+++}\,\{1\leftrightarrow2\} \, , \nn 
\AsqNLP{1524}{+-++}\,=\, \AsqNLP{1524}{-+++}\,\{1\leftrightarrow2\} \,,\nn 
\AsqNLP{1452}{+-++}\,=\, \AsqNLP{1245}{-+++}\,\{1\leftrightarrow2\} \,.
\end{align}
\item $ +++- $
\begin{align}
\AsqNLP{1245} {+++-}  & \,=\,  \bigg ( \frac{s_{12}}{ s_{15} s_{24}}-\frac{3}{ s_{15}}-\frac{3}{ s_{45}}+\frac{s_{24}}{s_{12} s_{45}}-\frac{s_{14} s_{25}}{ s_{12} s_{15} s_{45}} -\frac{s_{14} s_{25}}{s_{15} s_{24} s_{45}} \bigg )  \mathcal{A}^2 _{+++} \nn & \qquad+ \AsqNLP{1245} {++++} \,, \nn
\AsqNLP{1524} {+++-}  & \,=\, \AsqNLP{1245} {+++-} \,\{2\leftrightarrow4\}\,, \nn 
\AsqNLP{1452} {+++-} & \,=\, \AsqNLP{1245} {+++-} \,\{1\leftrightarrow2\} \,.
\label{eq:pppmNLPSq}
\end{align}
\end{enumerate}
Note that, the colour ordering of the non-radiative squared amplitude, suppressed here and in the rest of the paper, is to be considered as $\{124\}$ {\em i.e.,}  $\mathcal{A}^2_{h_1 h_2 h_4}=[\mathcal{A}^2]^{124}_{h_1 h_2 h_4}$. Each of the remaining five non-NMHV squared amplitudes resemble to one of the above results as their helicity amplitudes are obtained by flipping helicities of all the external particles. 

\subsection{Phase Space Integration}
We are now ready to integrate the squared amplitudes over the unobserved parton phase space in the rest frame of $ p_4 $ and $ p_5 $ momenta to obtain the differential cross-section. Following the usual method, we factorize the three-body phase space into two two-body phase spaces: ({\em i}) one containing two gluons with momenta $p_4$ and $p_5$, ({\em ii}) the other one containing the Higgs and the collective contribution of the two gluons mentioned in ({\em i}). We choose the phase space parametrisation in $ d=(4-2\epsilon)$ dimension~\cite{Ravindran:2002dc,Beenakker:1988bq} as, 
\begin{eqnarray}
	p_1&=&(E_1,0,\cdots,0,E_1) \,, \nonumber \\
	p_2&=& (E_2,0,\cdots,0,p_3\sin \psi, p_3\cos \psi - E_1) \,, \nonumber\\
	p_3&=&-(E_3,0,\cdots,0,p_3\sin \psi, p_3\cos \psi ) \,, \nonumber \\
	p_4&=&-\frac{\sqrt{s_{45}}}{2} (1,0,\cdots,0,\sin \theta_1 \sin \theta_2,\sin\theta_1\cos \theta_2,\cos \theta_1) \,, \nonumber\\
	p_5&=&-\frac{\sqrt{s_{45}}}{2} (1,0,\cdots,0,-\sin \theta_1\sin \theta_2,-\sin\theta_1\cos \theta_2,-\cos \theta_1)\,. 
	\label{eq:para1} 
\end{eqnarray}
The differential cross-section at NLP is then given by, 
\begin{align}
\left.	s_{12}^2\frac{d^2\sigma}{ds_{13}ds_{23}}\right|_{\text{NLP}}
	\,&= \, \mathcal{F} \left (\frac{s_{45}}{\mubar^2}\right )^{-\epsilon}\,\overline{\mathcal{A}_{\text{NLP}}^2} \,,
\end{align}
where
\begin{align}
	\mathcal{F}\,&=\, \frac{1}{2}K_{gg}\,G^2\left( \frac{\alpha_{s}(\mubar^2)}{4\pi}\right )^2\,
	\left(\frac{s_{13}\,s_{23}-m_H^2\,s_{45}}{\mubar^2\,s_{12}}\right )^{-\E} \,, \nn 
	K_{gg}\,&=\frac{N^2}{2(N^2-1)}, \quad \mubar^2= 4\pi e^ {-\gamma_{E}\epsilon} \mu_r^2 \,,
\end{align}
and
\begin{align}
	\overline{\mathcal{A}_\text{NLP}^2}\,=&\,\int_{0}^{\pi} d\theta_1\, (\sin\theta_1)^{1-2\E}\int_{0}^{\pi}d\theta_2\, (\sin\theta_2)^{-2\E} \AsqNLP{} \,.
\end{align}
We can now use eqs.~\eqref{eq:ppppNLPsq} --~\eqref{eq:pppmNLPSq} and formulae given in~\cite{Lyubovitskij:2021ges} to perform the angular integrations which give us the NLP threshold logarithms.
We have checked that the singular terms produced after these integrations due to the hard collinear emissions get cancelled, once the effects of mass factorization using helicity dependent Altareli-Parisi splitting functions~\cite{Altarelli:1977zs, Larkoski:2013yi} are taken into account. The helicity driven NLP leading logarithms that contribute to the differential cross-sections are given by,
\begin{enumerate}
	\item $ \displaystyle ++++  $
	\begin{align}
	\left.	s_{12}^2\frac{d^2\sigma_{++++}}{ds_{13}ds_{23}}\right|_{\text{NLP-LL}}
	\,=\,\mathcal{F}\, & \Bigg \{  
		16 \pi  \left(s_{12} \left(\frac{1}{s_{13}}+\frac{1}{s_{23}}\right)+2\right) \log \left (\frac{s_{45}}{\mubar^2}\right ) \nn & 
		+16 \pi  \log \left(\frac{s_{12}s_{45}}{s_{13}s_{23}} \right)\Bigg \} \times \frac{1}{\mhsq} \, \mathcal{A}^2 _{+++}\,.
		\label{eq:NLPlog1}
	\end{align}
	\item $ -+++ $
	\begin{align}
\left.	s_{12}^2\frac{d^2\sigma_{-+++}}{ds_{13}ds_{23}}\right|_{\text{NLP-LL}}
\,=\,\mathcal{F} \, \Bigg \{& 
		16 \pi  \left (\frac{1}{s_{13}}-\frac{1}{s_{23}}\right )\log \left ({\frac{s_{45}}{\mubar^2}}\right ) \nn &
		+4 \pi \left (\frac{3}{s_{13}} -\frac{1}{s_{23}} \right ) \log \left(\frac{s_{12} s_{45}}{s_{13} s_{23}} \right)
		\Bigg\} \, \mathcal{A} ^2_{-++} \,.
	\end{align}
	\item $ ++-+ $
	\begin{align}
\left.	s_{12}^2\frac{d^2\sigma_{++-+}}{ds_{13}ds_{23}}\right|_{\text{NLP-LL}}
\,=\, \mathcal{F} \Bigg \{ & 16 \pi \left (\frac{1}{s_{13}}+\frac{1}{s_{23}}\right ) \log \left ({\frac{s_{45}}{\mubar^2}}\right ) \nn & -4\pi \left (\frac{1}{s_{13}}+\frac{1}{s_{23}}\right ) \log \left(\frac{s_{12} s_{45}}{s_{13} s_{23}} \right) \Bigg \}  \, \mathcal{A}^2_{++-} \,.
	\end{align}
	\item $ +-++ $
	\begin{align}
		s_{12}^2 \left.	s_{12}^2\frac{d^2\sigma_{+-++}}{ds_{13}ds_{23}}\right|_{\text{NLP-LL}}
		\,=\, \mathcal{F} \bigg \{& 16 \pi \left (\frac{1} {s_{23}}-\frac{1}{s_{13}}\right )\log \left(\frac{s_{45}}{\mubar ^2}\right)\nn& +4 \pi \left (\frac{3  }{s_{23}}-\frac{1}{s_{13}}\right ) \log \left (\frac{s_{12}s_{45}}{s_{13}s_{23}}\right) \bigg\} \, \mathcal{A}^2_{+-+} \,.
	\end{align}
	\item $ +++- $
	\begin{align}
		\left.	s_{12}^2\frac{d^2\sigma_{+++-}}{ds_{13}ds_{23}}\right|_{\text{NLP-LL}}
		\,=\, \mathcal{F} \Bigg \{  & - 16 \pi \bigg (\frac{1}{s_{13}}+\frac{1}{s_{23}}\bigg)\log \left( \frac{s_{45}}{\mubar^2}\right ) \nn& - 4 \pi \left  (\frac{1}{s_{13}}   +\frac{1}{s_{23}}  \right ) \log \left (\frac{s_{12}s_{45}}{s_{13}s_{23}}\right)\Bigg\}\, \mathcal{A}^2_{+++}
		 \nn & + \left.	s_{12}^2\frac{d^2\sigma_{++++}}{ds_{13}ds_{23}}\right|_{\text{NLP-LL}} \,.
		\label{eq:NLPlog5}
	\end{align}
\end{enumerate}
From the above equations, it is evident that the threshold variable is $ \displaystyle \xi =\left(\frac{s_{45}}{\bar{\mu}^2}\right ) $.
Flipping of all helicities together in each one of the above equations does not alter the result. Therefore the complete result can be achieved by adding eqs. \eqref{eq:NLPlog1} -- \eqref{eq:NLPlog5} and then by multiplying a factor of $2$. 
\section{Summary and Outlook}
\label{sec:conclu}
The avalanche of high accuracy data in the LHC demands perturbative QCD predictions to be extremely precise. From theoretical point of view, all order resummation and fixed order calculations both are important to reach the desired precision. NLP corrections can leave numerically sizeable impact on the differential distribution of cross sections in the threshold limit. Although there exists a method to calculate NLP corrections using momentum shifts at the squared amplitude level, the rarity of results clearly demands improvement on the method of such calculations.  

We have considered the effect of next-to-soft radiation on the Higgs plus one jet production through gluon fusion. 
We have shifted the spinors in the non-radiative helicity amplitudes which essentially generate the helicity amplitudes in the case of an extra gluon emission in the next-to-soft limit. 
The squared amplitudes thus obtained are compact in nature and it comes out that the NMHV amplitudes do not play a role in the calculation of threshold logarithms. We have performed the phase space integration over the unobserved parton phase space to obtain the NLP threshold logarithms and listed the results for each helicity configurations. A systematic method to calculate NLP leading logarithms is presented in this paper exploiting the connection between the shifted dipole spinors and colour ordered radiative amplitudes, for the first time, at the helicity amplitude level. We believe that the simplicity and easy applicability of the approach presented here would facilitate bringing out more such results for several other processes.    
\section*{Acknowledgments}
We thank Keith Ellis and Eric Laenen for their useful comments on the manuscript.  SS is supported in part by the SERB-MATRICS under Grant No.
MTR/2022/000135.

\bibliographystyle{bibstyle}
\bibliography{ref.bib}

\end{document}